\documentclass[aps,%
twocolumn,
groupedaddress]{revtex4}
\input epsf
\usepackage{epsfig}

\begin{document}

\title{Cuts and penalties: a comment on 
``The clustering of ultra-high energy cosmic rays
  and their sources''}  
\author{P.G.~Tinyakov$^{a,c}$ and I.I.~Tkachev$^{b,c}$\\
$^a${\small\it Institute of Theoretical Physics, University of
Lausanne,} \\ {\small\it CH-1015 Lausanne, Switzerland}\\
$^b${\small\it CERN Theory Division, CH-1211 Geneva 23, Switzerland}\\
$^c${\small\it Institute for Nuclear Research, Moscow 117312, Russia }
}

\begin{abstract}
  In a series of papers we have found statistically significant
  correlations between arrival directions of ultra-high energy cosmic
  rays and BL Lacertae objects.  Recently, our calculations were
  partly repeated by Evans, Ferrer and Sarkar \cite{Evans:2002jy} with
  different conclusions.  We demonstrate that the criticism of
  Ref.~\cite{Evans:2002jy} is incorrect.  We also present the details
  of our method.
\end{abstract}


\pacs{PACS numbers: 98.70.Sa}

\maketitle

\section{Introduction}

Identification of sources of the ultra-high energy cosmic rays (UHECR)
is a key to understanding their nature. The observed small-scale
clustering of
UHECR \cite{Hayashida:bc,Takeda:1999sg,Tinyakov:2001ic,Takeda:2001xx}
suggests that already existing data
\cite{Takeda:1999sg,experiments,Hayashida:2000zr} may contain information sufficient
to make first steps in this direction. In a series of papers
\cite{Tinyakov:2001nr,TT_ICRC,Tinyakov:st,Tinyakov:2001ir,Gorbunov:2002hk,Tinyakov:2002nm,Tinyakov:2003nu}
we have shown that there exist significant correlations between the
arrival directions of UHECR and BL Lacertae objects (BL Lacs), and
therefore BL Lacs are likely to be sources of UHECR. Although present
small dataset does not allow to determine with certainty the details
of UHECR production and propagation, interesting hints may be obtained
\cite{Tinyakov:2001ir,Gorbunov:2002hk,Tinyakov:2002nm,Tinyakov:2003nu}.

The first evidence of connection between UHECR and BL Lacs was found
in Ref.~\cite{Tinyakov:2001nr} where we have shown that the combined
set of AGASA events with $E>4.8\times 10^{19}$~eV and Yakutsk events
with $E>2.4\times 10^{19}$~eV (identified previously as the sets with
largest clustering) correlates strongly with most powerful BL
Lacs. The statistical significance of this correlation was found to be
$6\times10^{-5}$ with penalty factor included (for explanation of
penalty factor see below).

In the recent paper by Evans, Ferrer and Sarkar (EFS)
\cite{Evans:2002jy} the validity of this result was called into
question.  The rational part of the criticism of EFS boils down to the
following two issues: i) choice of cuts on BL Lacs and/or calculation
of the penalty factor ii) choice of the cosmic ray set. Below we
address these issues and demonstrate that criticism of EFS is
unjustified. We also clarify some frequently arising questions related
to our analysis.

\section{Statistical significance of correlations: cuts and penalty factors}
\label{subsect:penalties}

\subsection{What is penalty factor} 

In statistical analysis, one nearly always has to make cuts in order
to improve signal-to-noise ratio. The question is how to take them
into account correctly. A common wisdom is that cuts should be fixed
{\it a priori}, i.e., based on theoretical considerations. In that
case the cuts simply limit the number of data points, but do not alter
the calculation of probabilities.

Not always this is possible. For instance, in the case of UHECRs,
their acceleration mechanism is not known. How to impose cuts in a
catalogue of astrophysical objects in order to select actual UHECR
emitters? In which band --- radio, optical, X-ray, $\gamma$, TeV? To
which flux limit? In Ref.~\cite{Tinyakov:2001nr} we proposed an
approach to this problem which consists in adjusting cut(s) so as {\it
to maximize} the signal, and then compensating this 
cut adjustment by a {\it penalty factor}.

A penalty factor takes into account the ``number of independent
attempts'' made when searching for the best signal. For instance, if
two independent catalogues were tried, the penalty factor is 2 (in the
limit of small probabilities). If $N$ catalogues tried are not
independent, as in the case of cuts in a single catalogue, the penalty
factor is smaller than $N$ and should be calculated by means of the
Monte-Carlo simulation.

In correlation analysis of cosmic rays, the quantity of interest is
the probability $p_{\rm min}^{\rm data}$ that the observed excess of
cosmic rays around source positions is the result of a chance
coincidence. This probability depends on cuts made in the source
catalogue; the latter are adjusted so that the probability is minimum
(correlations are maximum). The cut adjustment should be compensated
by the penalty factor which is calculated as follows. A random set of
cosmic rays is generated and treated exactly as the real data: the
same cuts are tried and the resulting minimum probability $p_{\rm
min}^{\rm MC}$ is determined.  This procedure is then repeated for a
large number of random ``cosmic ray'' sets, and $p_{\rm min}^{\rm MC}$
is determined each time. The number of occurrences of a value $p_{\rm
min}^{\rm MC} \leq p$ is then counted as a function of $p$. Divided by
the total number of sets, this gives the probability $P(p)$ that the
adjustments of cuts produces $p_{\rm min}\leq p$ for a random set of
cosmic rays. $P(p_{\rm min}^{\rm data})$ is the correct measure of the
significance of observed correlations.

When no adjustment of cuts is made one obviously has $P(p)=p$, and the
quantitative measure of correlation is the probability $p$
itself. When cuts are adjusted, small probabilities appear more often
by construction, and this relation is modified. The modification can
be written in terms of the penalty factor $F(p)$,
\begin{equation}
P(p) = F(p) \cdot p.
\label{eq:penaltydef}
\end{equation}
In the presence of cut adjustment, the significance of
correlations is determined by the product $F(p_{\rm min}^{\rm
data})\cdot p_{\rm min}^{\rm data}$.

To summarize, all parameters characterizing UHECR sources and UHECRs
themselves can be subject to cuts. Some of these cuts can be decided a
priori; they reflect (and depend on) physical assumptions made. These
cuts do not require penalty. One can call these cuts {\em
fixed}. Alternatively, in the absence of physical arguments, the cut
should be chosen (adjusted) so as to maximize the signal. These cuts
can be called {\em adjustable}. These are the adjustable cuts which
imply non-trivial penalty factors; the latter can be calculated in a
way explained above.

\subsection{UHECR set}

In our analysis \cite{Tinyakov:2001nr}, the cosmic ray set was fixed
on physical grounds using results of previous publications
\cite{Tinyakov:2001ic}, and was not adjusted in search for
correlations. The motivation was as follows. It was observed earlier
\cite{Hayashida:bc,Takeda:1999sg,Takeda:2001xx} that highest-energy
cosmic rays exhibit remarkable property: they auto-correlate on the
angular scale consistent with the angular resolution of detectors. The
mere existence of these correlations suggests that a relatively small
number of point sources of cosmic rays is contributing a considerable
fraction of the UHECR flux \cite{Dubovsky:2000gv}, and that the trace
of the original directions to sources is not lost completely. The
identification of these sources via cross-correlation analysis may
therefore be possible. Within the hypothesis that UHECR clusters are
due to point sources, the strongest correlation signal is expected
with the UHECR set which has most significant auto-correlations. The
latter requirement selects \cite{Tinyakov:2001ic} AGASA events with
$E>4.8\times 10^{19}$~eV and Yakutsk events with $E>2.4\times
10^{19}$~eV. The observed angular size of clusters also fixes the
angular scale at which correlations with sources are expected.  

The fact that AGASA and Yakutsk sets have different energy cuts in
this approach is, of course, disturbing.  It may, however, be
explained by different energy calibration of the two experiments, or
smaller number of the Yakutsk events. In either case, within our
assumptions it is inconsistent to change energy cuts or to drop
Yakutsk events once autocorrelations are found in this set.

The authors of Ref.~\cite{Evans:2002jy} have rejected the Yakutsk
events because Yakutsk array has angular resolution not as good as
AGASA. By itself, worse angular resolution does not imply that
correlations with sources {\em must} be absent in Yakutsk set: even
though the angular resolution is worse, the density of UHECR events
around actual sources is larger as compared to a random set, and one
has an excess in counts even at small angles. The actual presence of
autocorrelations in Yakutsk data supports this statement.  Therefore,
it is not correct that ``the correlations found at smaller
$\langle\mbox{than Yakutsk angular resolution} \rangle$ angles cannot
be meaningful'' \cite{Evans:2002jy}.

\subsection{Correlations with BL Lacs}

The authors of Ref.~\cite{Evans:2002jy} have essentially reproduced
our Monte-Carlo simulations: solid curve of Fig.~3 of
Ref.~\cite{Evans:2002jy} agrees with our calculations if Yakutsk
events are discarded. They concluded, however, that we have
underestimated the penalty factor. This conclusion was based on i) a
comparison of correlations in the case when one particular set of cuts
was chosen to the case when no cuts on BL Lacs were made at all ii)
smaller correlations of BL Lacs with a different set of UHECR. Neither
such a comparison, nor the dependence on the UHECR set have anything
to do with the penalty factor. The UHECR set was {\em fixed} as
described above and was not adjusted during the calculation, thus
requiring no penalty. The penalty associated with the adjustment of BL
Lac set should be {\it calculated} as described above. This
calculation was performed in our paper
\cite{Tinyakov:2001nr} as follows.

The most complete catalog of QSO \cite{veron1} contains 306 confirmed
BL Lacs for which the apparent magnitude, redshift (where known) and
6~cm radio flux are listed (for several objects the 11~cm radio flux
is also given). In Ref.~\cite{Tinyakov:2001nr} we considered BL Lacs
with $z>0.1$ as suggested by statistics of clustering. (We also
examined a complimetary small set without further cut adjustment. This
``independent attempt'' adds one to the penalty factor.) We adjusted
cuts on magnitude and 6~cm radio flux in the following way. First, the
grid of $11\times 11$ cuts was fixed. Namely, the cut on magnitude was
varied from $m<16$ to $m<20$ with the step $0.4$; simultaneously the
cut on 6 cm radio flux was changed from 0.21~Jy to 0.01~Jy with the
step 0.02~Jy. Note that further ``refining'' of this set of cuts would
not change the results as the sets of BL Lacs obtained would overlap
almost completely. The lowest probability (best correlation) found
\cite{Tinyakov:2001nr} in the real data in this way equals to $p_{\rm
min}^{\rm data} = 4\times 10^{-6}$
\footnote{EFS quote a different number, $2\times 10^{-5}$. This is
merely misunderstanding of our paper.}.

\begin{figure}
\begin{picture}(240,170)(0,0)
\epsfig{file=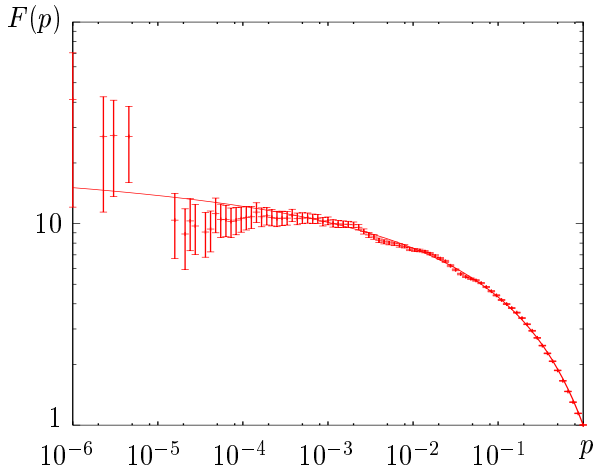,%
bbllx=114pt,bblly=360pt,%
bburx=360pt,bbury=535pt,%
width=230pt,height=180pt,%
clip=}
\end{picture}
\caption{MC calculation of the penalty factor $P(p)$.  Solid line
represents the fit by a power law in the log-log scale.}
\label{fig:penalty2}
\end{figure}

To calculate the penalty factor associated with cut adjustment, a
random configuration of 65 UHECR events was generated. Then its
correlation with all $11\times 11$ subsets of BL Lacs was examined and
the minimum probability $p_{\rm min}$ over $11\times 11$ cases was
found and recorded. The whole procedure was repeated for the next
random configuration.  A total of $N_{\rm tot}= 10^5$ random
configurations were treated in this way. The number of occurrences
$N(p)$ of $p_{\rm min}<p$ was calculated as a function of $p$. Divided
by $N_{\rm tot}$ it determines the corrected probability $P(p)$ and
the penalty factor $F(p)$ as defined in Eq.~(\ref{eq:penaltydef}). The
resulting penalty factor $F(p)$ is shown in Fig.~\ref{fig:penalty2}.
The error bars correspond to statistical errors in determination of
$N(p)$. 

Taking the penalty factor at the minimum probability obtained for the
real data, $F(4\times10^{-6})\simeq 14$ and adding 1 as penalty for BL
Lacs with $z<0.1$ as explained above, 
one finds the significance of correlations including penalty,
\[
P(4\times10^{-6})\simeq 6\times 10^{-5}.
\]
The significance is $\sim 4\sigma$ in terms of Gaussian distribution.  It
should be pointed out that, in order to get the correct significance, the
penalty factor should be multiplied by the {\it minimum} probability
obtained for the real data with the same set of cuts.

We stress again that the absence of correlations of cosmic rays with
the entire sample of known BL Lacs does not mean anything. Physically,
cosmic rays correlate with the brightest BL Lacs. Including dimmer BL
Lacs in the analysis dilutes the correlations, and this is the only
conclusion one draws from the analysis made in Ref.~\cite{Evans:2002jy}.

\subsection{``Correlations'' with GRB}
\label{sect:grb}

The authors of Ref.~\cite{Evans:2002jy} have chosen GRBs as a control
set where correlations are not expected; it has to be compared with
the case of BL Lacs. From the fact that correlations with the {\it
whole} catalogue are absent in both cases while there are many
coincidences, the authors of Ref.~\cite{Evans:2002jy} concluded that
GRBs correlate with UHECR ``just as well as do BL Lacs!'', implying
that correlations with BL Lacs are due to cut adjustment. This
conclusion can be tested by Monte-Carlo simulation: one has to check
whether cut adjustment leads to apparent correlations with GRBs. It
does not, as we now demonstrate.

The Fourth BATSE Gamma-Ray Burst Catalog \cite{Paciesas:1999tp}
contains 1637 objects from which we cut out, in sequential order
(according to observation date and time), 5 non-overlapping sets of
306 objects each (recall that 306 is the number of confirmed BL Lacs
in the catalogue \cite{veron1}). In each set we have looked for
maximum correlations by optimizing cuts. We present two different
tests: one-dimensional cut which allows transparent graphical
representation, and two-dimensional cut which mimics more precisely
the case of BL Lacs. We took the same maximally auto-correlated 
UHECR data set as in Refs. \cite{Tinyakov:2001nr,Evans:2002jy}.

\begin{figure}
\begin{center}
\begin{picture}(210,200)(10,0)
\put (0,0) {
\epsfig{file=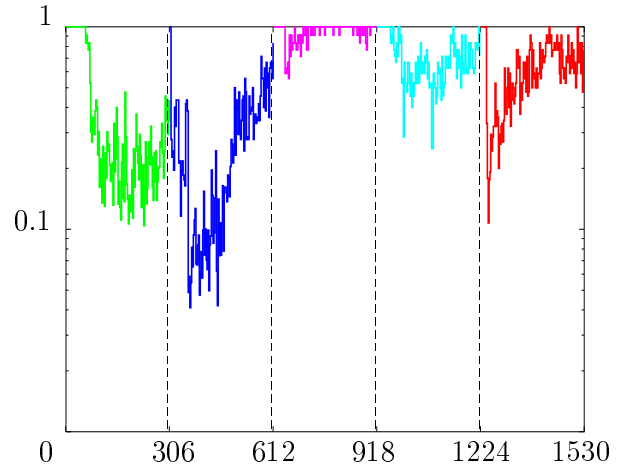,%
bbllx=95pt,bblly=330pt,%
bburx=338pt,bbury=515pt,%
width=230pt,height=180pt,%
clip=}}
\end{picture}
\end{center}
\caption{Correlations of UHECR with GRB: probability of the observed
excess of UHECR events near GRBs for 5 independent GRB subsets
as a function of the cut applied within each subset.}
\label{fig:GRB}
\end{figure}

In the first test we replace the $11\times 11$ cuts on magnitude and
radio flux by 121 cut on the total number of objects (we take them in
sequential order, so that first few cuts correspond to including 3, 6,
9 ... first objects, and the last cut corresponds to including all 306
objects)
\footnote{It is not important on which particular parameters the cuts
are imposed in the control set. A statement that some parameter is
{\it a priori} important is already equivalent to the statement that
correlations are present.}. Like in the case of BL Lacs, we then
calculate the probability of the observed excess of UHECRs around GRBs
as a function of the cut. The results are presented in
Fig.~\ref{fig:GRB}. The smallest probability found in all 5 sets is $>
4\%$ (which by itself is not statistically significant). In view of
the similar number of cuts in two cases, Fig. 1 provides an
estimate \footnote{The exact calculation of penalty factor is, of
course, possible along the lines of Sect.~IIA, but rough estimate is
sufficient in the case at hand.} for the penalty factor,
$F(0.04)\approx 5$, in a single set of 306 GRBs. Multiplying the best
probability, the penalty factor and the number of independent GRB sets
one gets a number of order 1.

To mimic the case of BL Lacs we perform a similar test with {\em two}
independent cuts. The first cut is done as before on the sequential
number of the event. The second cut is imposed on the time (in seconds
of the day, UT) of the event. Note that these two cuts are virtually
uncorrelated since BATSE registered roughly one event per day. We have
searched for the best correlation signal on the grid of $11\times 11$
equidistant cuts from 1 to 306 and from 1 to 86400, respectively. In
the above five catalogs consisting of 306 GRBs each, the following
lowest probabilities were found: 9.5\%, 1.4\%, 20\%, 1.8\%, 24\%.
Multiplying the lowest of these probabilities 0.014 by the penalty
factor $F(0.01)\approx 8$ and by the number of independent GRB sets,
we again obtain a number of order 1.

These examples are a good illustration of compensation between the
effects due to cut adjustment and the penalty factor.  When
compensated by the penalty factor, cut selection does not introduce
apparent correlations when in reality they are absent.

To avoid confusion, we note again that we used GRBs here as a control
set, and imposed physically unmotivated cuts on purpose.

\section{Subtleties of correlation analysis}

In this section we discuss some subtleties of correlation analysis
which were not mentioned in Ref. \cite{Evans:2002jy}.  We hope that this
will answer a number frequently arising questions regarding our work
and correlation analysis in general.

\subsection{Why completeness of the BL Lac catalogue is not necessary for 
establishing correlations with UHECR}
\label{sect:completeness}

A catalogue of astrophysical objects is called complete when it is
believed to contain more than, say, 90\% of all existing objects of a
given type, in a given region of the sky, and down to some fixed level
of luminosity in a given waveband(s). The catalogue \cite{veron1},
used in our analysis, is not complete simply because it contains all
objects known up to date without further selection \cite{veron2}.

It is a common wisdom that completeness of the catalogue is crucial
for statistical analysis. The reason is obvious: in an incomplete
catalogue, the distribution of objects in brightness, position, etc.,
reflects not only their actual abundance, but also observational
bias. When one studies, for instance, the evolution of abundance of
some objects with time, one has to make sure that logN/logS dependence
reflects actual change in spatial density rather than the fact that
remote (and therefore dim) objects are more difficult to
observe. Indeed, as it is, the catalogue \cite{veron1} is not suitable
for any analysis where {\it statistical properties of BL Lacs} are of
interest.

In the correlation analysis of UHECR the question is different: given
a set of candidate sources (BL Lacs in the case at hand), is the
distribution of cosmic rays random, or does it peak around BL Lac
positions? Clearly, these are the {\it statistical properties of
cosmic rays} which are of interest in that case, not the statistical
properties of BL Lacs. Technically speaking, when calculating the
correlation function by the algorithm of Ref.~\cite{Tinyakov:2001nr},
the cosmic ray directions are simulated, while source positions are
held fixed. For this reason the method is applicable without change
to studying, for instance, a correlation of UHECR with one particular
direction (say, the Galactic center), i.e. in the case when the notion
of completeness does not apply.

Completeness is not necessary for {\it establishing the fact} of
correlations: if statistically significant correlations are found with
an incomplete catalogue, this is a real signal. Incompleteness of the
catalogue of BL Lacs cannot be a source of correlations with
UHECR. Indeed, the objects which have to be added (removed) from a
catalogue to make it complete are absent from (present in) the
catalogue for reasons not related to cosmic rays. Therefore,
correlations with UHECR can only be weakened by such an
incompleteness.

\subsection{UHECR autocorrelations}
\label{sect:auto}

The UHECR set used in our calculations is known to contain event
clusters --- in fact, it was chosen in such a way that
auto-correlations are maximum. When studying cross-correlations of
such a set with potential sources, one has to be careful to take
clusters into account when calculating the probability of chance
coincidence. Namely, one has to make sure that the signal observed in
the data is not due to chance coincidence of {\em clusters} of cosmic
ray events with candidate sources. For a given cluster, such a
coincidence is roughly as probable as for a given single
event, but contributes more into correlation function. This could
produce artificial enhancement of correlations if not taken into
account in the Monte-Carlo simulation.

In our calculations \cite{Tinyakov:2001nr} this problem is solved by
introducing in each Monte-Carlo cosmic ray set the same number of
doublets and triplets as there are in the real data. Then {\em chance}
coincidences between clusters and candidate sources happen in
simulated sets as often as in the real data, and thus are correctly
accounted for. This is confirmed by calculations presented in
Sect.~\ref{sect:grb} where correlations with GRB catalogs do not
appear despite auto-correlations in the UHECR dataset.

\subsection{Choice of angular scale}
\label{sect:angularScale}

The observed angular size of clusters suggests the angular scale at
which correlations with sources are expected. In our analysis
\cite{Tinyakov:2001nr} we have fixed it to the previously published
\cite{Hayashida:bc,Takeda:1999sg} AGASA value of $2.5^{\circ}$. 
This value is treated as not adjustable.

When the angular resolution of the experiment is known, there exists a
preferred choice of the angular scale --- the scale at which
correlations are {\em expected} to be maximum. This angle can be
determined by means of the Monte-Carlo simulation as follows. One has
to generate cosmic ray sets which are correlated with sources (taking
into account the experimental angular resolution) and then measure
angle at which the correlation signal is maximum. In the case of AGASA
events this procedure gives the result very close to $2.5^{\circ}$
(see dotted curve on Fig.~3 of Ref.~\cite{Tinyakov:2001nr}), so in
fact the choice of Ref.~\cite{Tinyakov:2001nr} was close to optimal.

Alternatively, one may choose not to fix the angular scale and treat
it as a free parameter. Then one should adjust it to maximize the
correlation and calculate corresponding penalty factor. A serious
problem of this approach is that the result would depend on the limits
within which the angular scale is varied. So finally one would have to
input in one or the other way the information about expected angular
scale of correlations in order to obtain a definite answer. 
We do not follow this approach in our calculations.

\subsection{Hidden penalty}
\label{sect:hidden}

We believe that we have accounted for all contributions to the penalty
factor (``effective number of tries'') {\it directly related} to our work.
Still, the obtained significance may be somewhat overestimated. The
point is that we are not the first who are looking for correlations
between UHECR and astrophysical objects. Some of these attempts have
been published in the literature; the others may have never been
reported. All these attempts should contribute, in principle, into the
penalty factor. However, it does not seem possible to account
correctly for all such contributions.

The way around this problem is obvious. First, only very low values of
$P$ should be interpreted as a signal (in our calculations we
considered $P < 10^{-4}$ to be sufficiently low to report our
results). Second, and more important, the results have to be confirmed
with a new independent data set.

\section{Further evidence}
\label{Further}

The correlations found in Ref.~\cite{Tinyakov:st} for a particular set
of UHECR and BL Lacs suggest, strictly speaking, only that UHECR and BL
Lacs are connected to each other. They say little about the
acceleration mechanism, particle nature and other relevant physical
parameters. These questions can be fully addressed only with much
larger dataset than available now. Some attempts, however, can be
made. The final purpose is to arrive at the end at a consistent
picture which incorporates all known features of UHECR.  Here is a
sketch of these attempts, each bringing additional evidence of the
connection between UHECR and BL Lacs.

In Ref.~\cite{Gorbunov:2002hk} it was noted that cuts in BL Lac
catalogue, chosen in Ref.~\cite{Tinyakov:2001nr} so as to maximize
correlations with UHECR, select automatically $\gamma$-ray loud BL
Lacs. When this observation is consistently elaborated, i.e. a
sub-sample of the catalogue \cite{veron2} is selected on the basis of a
single criterium, namely, the cross-correlation with the EGRET
sources, the resulting subset of 14 $\gamma$-ray loud BL Lacs
correlates with UHECR at the level of $10^{-7}$ of chance
coincidence. This number cannot, of course, be interpreted as the
significance of correlations between UHECR and BL Lacs because of {\it
a posteriori} selection; rather the conclusion is that $\gamma$-ray
loudness may be a distinctive feature of those BL Lacs which are UHECR
accelerators.

In Ref.~\cite{Tinyakov:2001ir} an attempt was made to determine the
charge composition of UHECR by reconstructing actual arrival
directions of UHECR particles bent in the Galactic magnetic field. The
idea was that such a reconstruction should improve correlations of
UHECR with BL Lacs if the latter are the sources. Substantial
improvement was indeed observed for particles with the charge +1,
which is an indication of the presence of protons.

In Ref.~\cite{Tinyakov:2002nm} it was observed that if correlations of
UHECR and BL Lacs were due to chance coincidence, the coinciding rays
would be distributed over the sky randomly, reflecting only the local
density of BL Lacs and exposure of a cosmic ray experiment.  Thus, any
significant deviation in the distribution of {\it correlating} rays
over the sky from this expectation speaks in favor of real physical
connection between cosmic rays and BL Lacs. In fact, the UHECRs
correlating with BL Lacs form two ``spots'', with low probability to
occur by chance \cite{Tinyakov:2002nm}. This non-uniformity of the
distribution of correlating rays may be due to several factors: (1)
anisotropy of extragalactic magnetic fields at scales of order
$500$~Mpc; (2) poor knowledge of the Galactic magnetic field in some
areas of the sky; (3) fluctuations in the space distribution of the
nearest sources.

\section{Conclusions}
\label{sect:conclusions}

According to textbooks, any successful statistical analysis consists
in formulation of a ``null hypothesis'' and its subsequent {\it
falsification}, at some confidence level, by comparing to the
experimental data.  In the case at hand the ``null hypothesis'' which
is being tested is that BL Lacs (and, in particular, any subset of
them) and UHECR are uncorrelated. It takes {\it only one}
counter-example to disprove a hypothesis, while ``pro-examples'' {\it
do not} prove its validity. An illustration of this general rule has
been discussed in Sect.~II: the absence of significant correlations
of UHECR with the whole BL Lac catalogue {\it does not prove} that
UHECR and BL Lacs are uncorrelated (i.e., that there is no subset
of BL Lacs which are sources of UHECR and thus correlate with them).

Having found such counter-example (i.e. the case when correlation is
significant), the only thing which can be concluded, to a certain
confidence level, is that UHECR and a particular subset of BL Lacs are
correlated. The nature and physical implications of these correlations
have to be studied separately by formulating and testing different
``null
hypothesis''. Refs.\cite{Tinyakov:2001ir,Gorbunov:2002hk,Tinyakov:2002nm}
are first attempts in this direction.

\acknowledgements
 
{\tolerance=400 We are grateful to V.~Rubakov, D.~Semikoz,
M.~Shaposhnikov and P.~Veron for reading the manuscript and valuable
comments.  The work of P.T. is supported in part by the Swiss Science
Foundation, grant 20-67958.02. }

\appendix

\section{Statistics of clustering and the number of UHECR sources.}

The authors of Ref.~\cite{Evans:2002jy} have misinterpreted our
earlier paper \cite{Dubovsky:2000gv} which concerns the statistics of
clustering of UHECR and the number of their sources. In view of the
growing confusion we would like to clarify this issue.

In Ref.~\cite{Evans:2002jy} one reads: ``{\rm $\langle$observed
occurrence of clusters of UHECR$\rangle$} \dots was used to estimate the
spatial density of sources to be $6\times 10^{-3}\; {\rm Mpc}^{-3}$
\cite{Dubovsky:2000gv}. This would obviously place stringent
constraints 
on candidate astrophysical sources, e.g. $\gamma$-ray bursts (GRBs)
have a spatial density of only $\sim 10^{-5}\;{\rm Mpc}^{-3}$. 
However, a more careful analysis \cite{Fodor:2001yi}
shows that the uncertainties in this estimate are very large. The true
number is $n = 2.77^{+96.1(916)}_{-2.53(2.70)} \times 10^{-3}\; {\rm
Mpc}^{-3}$ at the 68\% (95\%) c.l.; moreover relaxing the assumptions
made, viz. that the sources all have the same luminosity and a
spectrum $\propto E^{-2}$, increases the allowed range even further,
e.g. to $n = 180^{+2730(8817)}_{-165(174)} \times 10^{-3}\; {\rm
Mpc}^{-3}$ \dots''. 

The actual situation is different from what is described in this
extract. First, intrinsic inaccuracy of the estimate was fully
acknowledged in Ref.~\cite{Dubovsky:2000gv}. Its physical reason is
obvious: clustering is not sensitive to the number of dim sources. The
estimates of Ref.~\cite{Fodor:2001yi} cited above are a good
illustration of the point: {\it upper} limits are huge and strongly
model-dependent. 

Second, the main point of Ref.~\cite{Dubovsky:2000gv} was to show that
there exists a {\it model-independent lower bound} on the number of
sources. This bound is presented in Table~1 of
Ref.~\cite{Dubovsky:2000gv}: at 90\% and 99\% confidence levels the
number of sources is $n>2.3\times 10^{-4}\;{\rm Mpc}^{-3}$ and
$n>3.2\times 10^{-5}\;{\rm Mpc}^{-3}$, respectively. This is consistent
with the calculations of Ref.~\cite{Fodor:2001yi} performed in
particular models.

\end{document}